# Metasurface spectrometers beyond resolution-sensitivity constraints

Feng Tang[1,2†], Jingjun Wu[1,5†], Tom Albrow-Owen[3†], Hanxiao Cui[4†], Fujia Chen[2,8], Yaqi Shi[2,8], Lan Zou[6], Jun Chen[1], Xuhan Guo[7], Yijun Sun[2,8], Jikui Luo[2,8], Bingfeng Ju[8], Jing Huang[6], Shuangli Liu[6], Bo Li[1], Liming Yang[5], Eric Anthony Munro[3], Wanguo Zheng[5], Hannah J. Joyce[3], Hongsheng Chen[2,8], Lufeng Che[2], Shurong Dong[2,8], Tawfique Hasan[3*], Xin Ye[1,6*], Yihao Yang[2,8*], Zongyin Yang[2,8*]

[1] Mianyang Sci-Tech City Institute of Photon Technology; Mianyang, 621025, China.

[2] College of Information Science and Electronic Engineering, Zhejiang University; Hangzhou, 310027, China.

[3] Department of Engineering, University of Cambridge; Cambridge, CB3 0FA, UK.

[4] School of Aeronautics and Astronautics, Sichuan University; Chengdu, 610065, China

[5] School of Electronic and Optical Engineering, Nanjing University of Science and Technology; Nanjing, 210094, China.

[6] School of Information Engineering and School of Life Science and Engineering, Southwest University of Science and Technology; Mianyang, 621010, China.

[7] Department of Electronic Engineering, Shanghai Jiao Tong University; Shanghai, 200240, China.

[8] ZJU-Hangzhou Global Science and Technology Innovation Center and ZJU-UIUC Institute, Zhejiang University; Hangzhou, 310027, China.

*Corresponding author. Email: yangzongyin@zju.edu.cn; xinye@swust.edu.cn; th270@cam.ac.uk; yangyihao@zju.edu.cn

†These authors contributed equally to this work.

**Optical spectroscopy plays an essential role across scientific research and industry for non-contact materials analysis[1-3], increasingly through in-situ or portable platforms[4-6]. However, when considering low-light-level applications, conventional spectrometer designs necessitate a compromise between their resolution and sensitivity[7,8], especially as device and detector dimensions are scaled down. Here, we report on a miniaturizable spectrometer platform where light throughput onto the detector is instead enhanced as the resolution is increased. This planar, CMOS-compatible platform is based around metasurface encoders designed to exhibit photonic bound states in the continuum[9], where operational range can be altered or extended simply through adjusting geometric parameters. This system can enhance photon collection efficiency by up to two orders of magnitude versus conventional designs; we demonstrate this sensitivity advantage through ultra-low-intensity fluorescent and astrophotonic spectroscopy. This work represents a step forward for the practical utility of spectrometers, affording a route to integrated, chip-based devices that maintain high resolution and SNR without requiring prohibitively long integration times.**

A central aim in optical device design is to make the most efficient use of photons available, and in doing so, extract as much information as possible from incident radiation. This is especially the case when looking to accurately characterize low-intensity light spectra; the efficiency or sensitivity of a system dictates a system's capability to collect data with satisfactory resolution in a given timeframe [10]. However, a major constraint in spectrometer design – in particular for miniaturized systems – lies in an inherent trade-off between the resolution and sensitivity of devices. For platforms based on conventional strategies (e.g. those using grating-mediated dispersion, or filter arrays), higher resolutions have typically had to come at the expense of the transmitted light intensity incident onto the detectors, leading to decreasing signal to noise ratios (SNRs) and longer integration times[7,8]. This limitation can be described *via* the resolution-luminosity product[7], $E = RL$, (where $R$ is the resolution and $L$ is luminosity, the light throughput from the source onto the detectors), given that the device's efficiency, $E$, is constant for a given sensing area. Currently the most feasible option for moving beyond this constraint is to use cooled, high-sensitive detectors[11], an impractical solution for many in-situ or portable devices.

Here, we report on a planar microspectrometer platform based around dielectric metasurface encoders which overcomes these limitations by exploiting photonic quasi-bound states in the continuum (qBICs). These encoders, combined with a computational reconstruction algorithm[4,12-14], enable a platform where detector flux in the device increases with resolution, affording a significantly more efficient system, where high sensitivity and performance can be simultaneously realized. Through excitation intensity-resolved fluorescence spectroscopy of bacteria samples, as well as telescopic planetary spectroscopy under different visibility conditions, we demonstrate the strengths of this system in being able to perform sensitive analyses under ultra-low irradiance, all while in a highly miniaturized form factor. These microspectrometers could afford significant advances for lab-on-a-chip-, drone- or satellite-based detection of subtle signals in applications such as trace chemical analysis, nanoscale biomedical sensing[10], or astrophotonics[15]. Furthermore, the planar, CMOS-compatible nature of the platform, requiring only a single lithographic patterning step, would allow for straightforward manufacture of a device array onto a CMOS imaging sensor. This could offer a route toward highly sensitive snapshot spectral imaging cameras which do not require any complex optics or moving parts.

**Bandstop strategy**

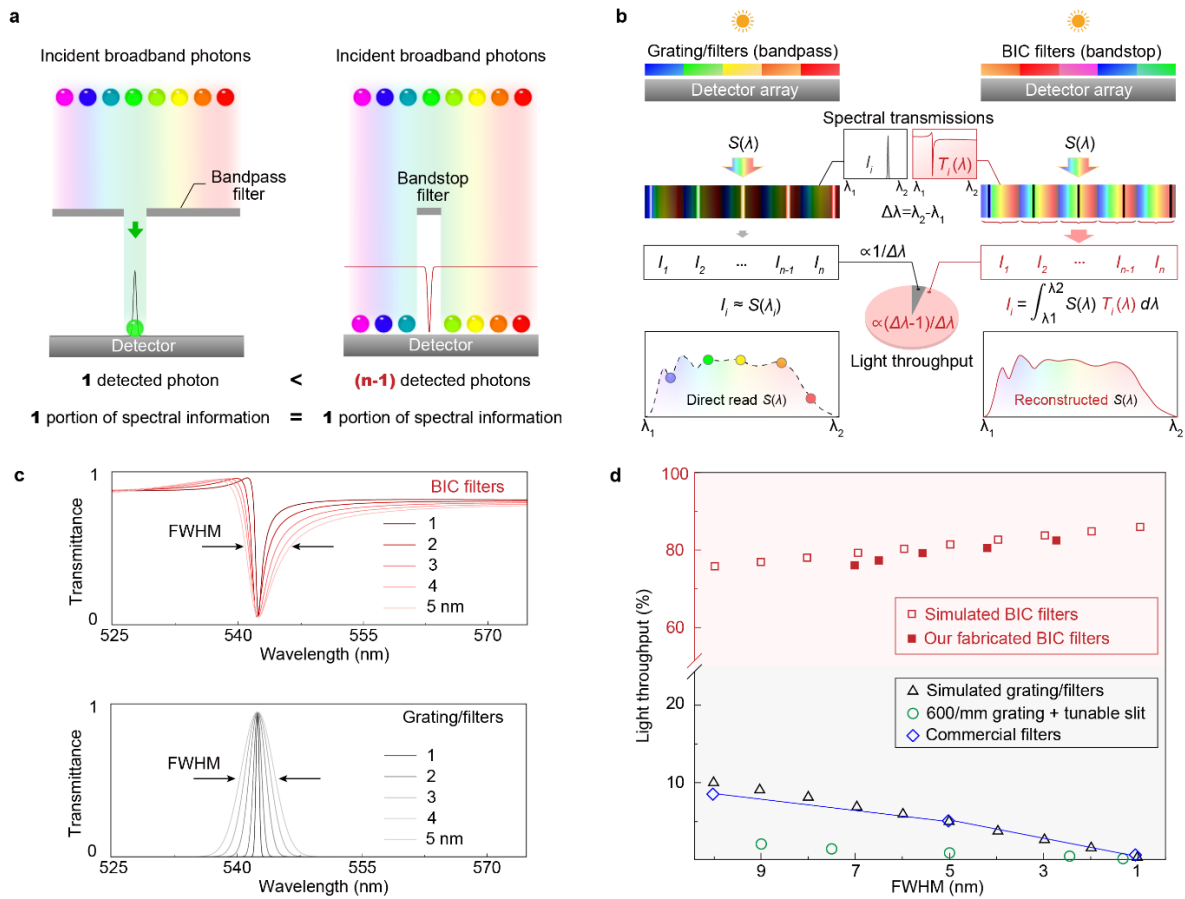

**Fig. 1 | Bandstop strategy for BIC-inspired spectrometer design. a,b** Comparison of bandpass and bandstop strategies when designing a filter-array based spectrometer. **c,** Simulated bandpass and qBIC bandstop features. **d,** Light throughput as a function of the FWHM of transmission features under illumination of a broadband light (500-600 nm), as transmitted through simulated and commercial grating and filter based systems with varying values of transmission FWHM.

The schematic in Fig. 1a illustrates the comparison of efficiency between bandpass and bandstop strategies. A bandstop system can transmit the same portion of spectral information while losing fewer photons. Figure 1b shows the differences in operational principle between our qBIC bandstop array spectrometer and conventional bandpass grating or filter array systems. In the latter, each detector reads light with a single wavelength component that comes from gratings or narrow bandpass filters. The number of filters, and their characteristic FWHM, dictate the effective resolution that such a system can achieve. Intuitively, given the nature of a bandpass transmission profile, where light outside of a target wavelength component is blocked, higher resolution (a narrower full-width at half maximum (FWHM) for each filter), necessarily results in lower flux and sensitivity.

A qBIC filter instead transmits the inverse, reflecting a narrow band of wavelengths. The characteristic transmission spectrum of a qBIC metasurface contains a collapsing Fano feature, resembling a "bandstop" profile[16,17] as seen in Fig. 1c. We integrate a range of different qBIC metasurfaces onto a CMOS detector array, to create a computational spectrometer device, in which wavelength components are not directly read, but where the qBICs are used to encode spectral information into the electronic responses of the sensors beneath them. An algorithm is developed to then reconstruct the spectra of incident light by solving the inverse problem posed by these bandstop encoders. Crucially, we show that the light throughput in a qBIC-based system is not only at least an order of magnitude higher than that of a bandpass system for broadband light, but the intensity is also proportional to the resolution or FWHM of the spectral feature (See Supplementary Information Section 1), rather than inversely proportional as seen in Fig. 1d. Such a qBIC-based s offers a route toward highly sensitive yet high-resolution micro-spectrometers.

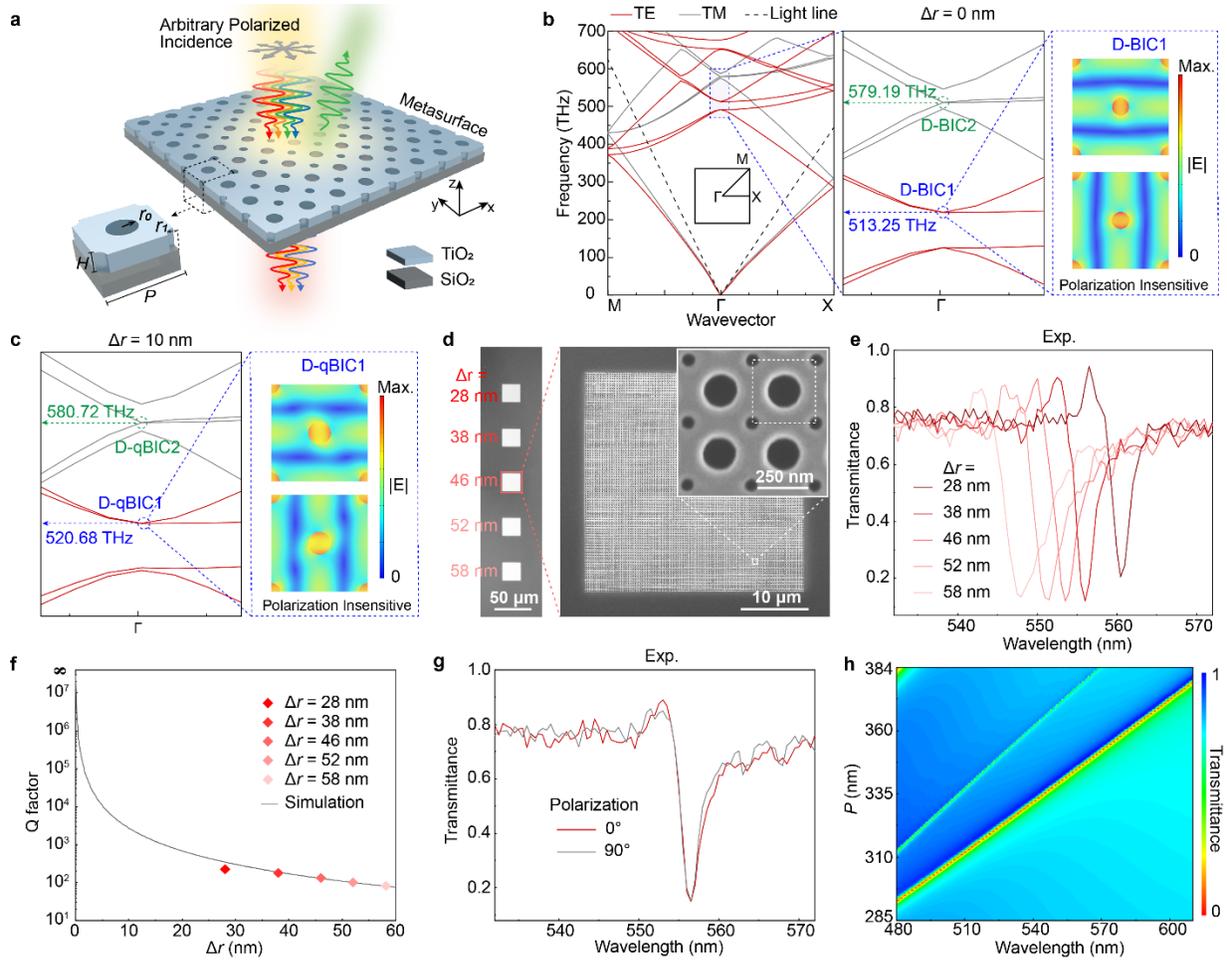

**Fig. 2 | Pixelated metasurfaces with polarization-independent qBICs. a,** Schematic diagram of the as-designed diatomic metasurface, formed from cylindrical nanoholes, arranged in a square lattice, etched into a TiO$_2$ slab with thickness $H$ = 92 nm. Unit cells (left) spaced at a pitch $P$, comprise of diagonally arranged holes with two different radii, $r_0$ and $r_1$. **b,** Band structures and field distributions of such a metasurface when $\Delta r = r_0 - r_1 = 0$ nm. The middle panel shows a magnified region of the band structure near the doubly-degenerate BICs. The right panel shows the field distributions of doubly-degenerate-BIC1. **c,** Band structure near doubly-degenerate-quasi-BIC (D-qBIC), and corresponding field distributions of D-qBIC1 when $\Delta r = 10$ nm. **d,** SEM images of a fabricated 1 × 5 qBIC-based metasurface filter array with $\Delta r$ = 28 nm, 38 nm, 46 nm, 52 nm and 58 nm, respectively. **e,** Measured transmitted spectra at varying values of $\Delta r$. **f,** Q factor of D-qBIC1 extracted from simulated and measured (**e**) transmitted spectra. **g,** Measured transmission spectra under 0 and 90-degree polarizations, demonstrating a polarization-independent response. **h,** Simulated metasurface transmission spectra for one hundred filters with varying unit cell pitch, $P$. For all other demonstrations in this figure, $P$ = 338 nm.

**Bound states in the continuum**

We first present simulations and experimentally demonstrate the physics of the qBIC encoders which underpin our platform. BICs are localized states with an infinitely long lifetime in the radiation continuum[9]; they usually manifest in the form of quasi-bound states[18], which can couple to the surrounding environment with dramatically enhanced light-matter interaction. They have demonstrated potential in a range of applications, such as low-threshold micro-/nano-lasers[19], high-sensitivity biosensors[16,17] and ultrafast vortex beam generators[20], though have yet to be applied toward spectroscopy. Figure 2a shows a schematic of the dielectric qBIC metasurface designed for this work, consisting of an array of cylindrical nanoholes arranged in a square lattice. Each diatomic unit cell of the metasurface includes two nanoholes – with radius $r_0$ and $r_1$. Unit cells are spaced at a pitch, $P$. Crucially for our devices, we will show that $P$ can be varied to tune the central wavelength of the metasurface's bandstop transmission feature, while $\Delta r = r_0 - r_1$ can be tuned to vary the FWHM of the transmission feature. The nanoholes are etched into a titanium dioxide ($TiO_2$) film deposited onto quartz at a thickness $H = 92$ nm. Here, $TiO_2$ provides low absorption whilst retaining sufficient index contrast with the quartz substrate (See Methods). When $r_0 = r_1 = 30$ nm, due to the band folding mechanism and the C4v symmetry of the metasurface[21,22], simulations show that the metasurface possesses doubly-degenerate BICs (D-BICs) at Γ, which can be decomposed into two orthogonal dipole-like modes that are a 90º rotation of each other (see Fig. 2b). In our metasurface design, there are two pairs of D-BICs: a transverse-electric (TE)-like mode at 513.25 THz (D-BIC1) and a transverse-magnetic (TM)-like mode at 579.19 THz (D-BIC2), respectively. They have similar physical characteristics and origins, and so for simplicity we focus here on analysis of D-BIC1 (See Methods). We introduce variations in $r_0$ so that $\Delta r$ is nonzero; this geometrical perturbation breaks the original translation symmetry but preserves the

C4v symmetry of the metasurface. In this way, the D-BICs degrade to doubly-degenerate quasi-BICs (D-qBICs) that are slightly coupled to the radiation continuum (see Fig. 2c) and which have finite but large Q factors (See Methods).

To experimentally confirm the characteristics of these D-qBICs, we fabricated five metasurfaces with $\Delta r$ = 28 nm, 38 nm, 46 nm, 52 nm and 58 nm, respectively. A full simulation of the relationship between metasurface transmission and the geometrical perturbation, $\Delta r$, is shown in the supplementary (See Supplementary Information Section 2). Figure 2d shows scanning electron microscopy (SEM) images of the metasurface slab array. Q factor values derived from both numerically simulated and experimentally measured (Fig. 2e) transmission spectra are displayed in Fig. 2f. There is an inverse quadratic dependence between the Q factor and $\Delta r$, and the Q factor tends to infinity when $\Delta r = 0$, as commonly seen in qBICs[18]. The experimental results are in good agreement with the simulated counterparts. For $\Delta r <= 40$ nm, the reduction in actual Q factor relative to simulations can be attributed to linewidth-broadening of the transmitted spectra due to measurement limitations imposed by the monochromator's resolution and SNR. Moreover, because of the preservation of C4v symmetry, the D-qBICs have been designed to be equally sensitive to light regardless of polarization (Fig. 2g)[21,22] and, as such, do not require a polarizer to filter the input light, further increasing the throughput of photons to the detector array. Crucially, we show through simulation the transmissive properties of D-qBICs as a function of $P$ (while maintaining fixed $\Delta r$, Fig. 2h); confirming that D-qBICs can function as tunable and narrow bandstop filters through modification of the $P$ value, enabling their use for spectral sensing across the visible range.

## qBIC spectrometers

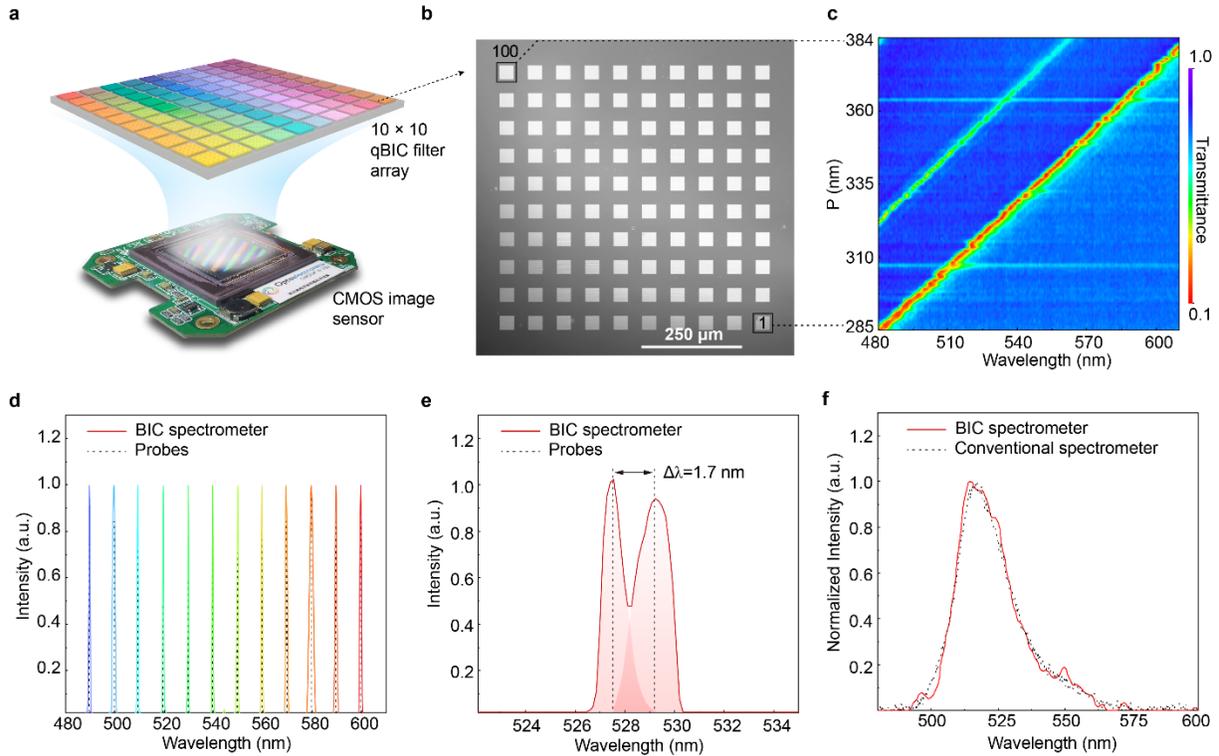

**Fig. 3 | Micro-spectrometer based on qBIC metasurfaces. a,** Schematic of the spectrometer featuring an individual 10 × 10 qBIC bandstop filter array integrated atop a CMOS imaging sensor. **b,** SEM image of the fabricated metasurface filter array. **c,** Bandstop transmission profiles $t_i(\lambda)$ of the 100 filters. **d,** Measured, reconstructed spectra of a series of monochromatic spectral lines (dark dotted lines) across the visible range. **e,** Measured, reconstructed spectrum of two narrow spectral lines, demonstrating ability to resolve distinct monochromatic spectral features down to 1.7 nm separation. **f,** Measured, reconstructed spectrum of a broadband light source, relative to the same incident signal measured by a conventional spectrometer.

Next, we demonstrate a high-performance micro-spectrometer based on our qBIC metasurface filters. Instead of a direct read-out, our spectrometer uses an array of qBIC metasurfaces, mounted on to a CMOS image sensor (Fig. 3a), to sample a light signal before computationally reconstructing the spectra. Represented mathematically, the light intensity incident on each of the detectors $I_i$ is a convolution of the incident spectrum $S(\lambda)$ and the transmitted profile $T_i(\lambda)$, given by a system of linear equations[4,12]:

$$\int_{\lambda_1}^{\lambda_2} S(\lambda)T_i(\lambda)\,d\lambda = I_i \qquad (i = 1,2,3\ldots,m) \qquad (1)$$

where $m$ is the detector number. By solving the system of equations gathered by all detectors via a reconstruction algorithm (See Methods), which also works to minimize the effects of measurement noise, we can arrive at an approximation of $S(\lambda)$.

The device is composed of a $10 \times 10$ metasurface filter array (Fig. 3b) with $P$-values vary linearly between 285 nm and 384 nm across the 100 filters, and the operational wavelengths of the corresponding metasurface filters across 480 – 610 nm. It is important to note that the operational wavelengths can be extended to other spectral regions simply via adjustments in both $P$ and $\Delta r$. The metasurface filter array is designed with a fix $\Delta r$ at 46 nm. This is a compromise between lowering $\Delta r$ to achieve higher resolutions, while being constrained by the limits of the transmission function calibration system and the noise level of the CMOS sensor array, as discussed in the supplementary (See Supplementary Information Section 3).

As shown in Fig. 3c, the transmission profiles $t_i(\lambda)$ of the metasurface filter array are measured under a customized micro-area UV-VIS spectrophotometer and show good agreement with the simulated responses in Fig. 2h. We note that the secondary bandstop feature seen at lower wavelengths does not interfere with the function of the spectrometer, as it is accounted for in the calibration of the system. Measurement of the spectral response profiles for the system as a whole, $T_i(\lambda)$ – a combination of the filters' transmission profiles $t_i(\lambda)$ and the photoresponse profiles $R_i(\lambda)$ of the CMOS pixels – is discussed in the supplementary (See Methods). While operating, under illumination by an arbitrary light signal with spectra $S(\lambda)$, the CMOS image sensor measures the set of light intensities, $I_i$, transmitted through the metasurface filter array. As described earlier in equation 1, the $I_i$ measured at each detector represents a set of linear equations, integrals of the incident spectrum $S(\lambda)$ and $T_i(\lambda)$ over the wavelength range; $S(\lambda)$ can be reconstructed by solving the inverse problem posed by this dataset. In practice, such a calculation is ill-posed, and highly susceptible to experimental noise; as such, an algorithm was developed to solve the set of linear equations generated by the encoders, the details of which are discussed in the supplementary text (See Methods).

To demonstrate the accuracy of the device, we tested the reconstruction fidelity of the microspectrometer when resolving a series of narrow spectral lines across a range of 480 – 610 nm. The results by the spectrometer match the positions of the individual probe lines with a mean accuracy of ± 0.11 nm and an average SNR of ~300, Fig. 3d. The dynamic range of spectrometer is ~500, calculated as the ratio of the detector saturation level and the lowest detectable intensity. For evaluating the spectral resolution of the device, we use the Rayleigh criterion: that two spectral lines are just resolved when the maxima of the first peak is positioned at the base of the second. As shown in Fig. 3e, this device enables a spectral resolution of 1.7 nm; such a resolution is comparable to that of most microspectrometer counterparts[4,5] (for reference, the resolution of a mature product, the Hamamatsu C12666MA microspectrometer, is 12 nm[23]). Our metasurface spectrometer can also accurately reconstruct continuous broadband spectra across the operational wavelength span as shown in Fig. 3f.

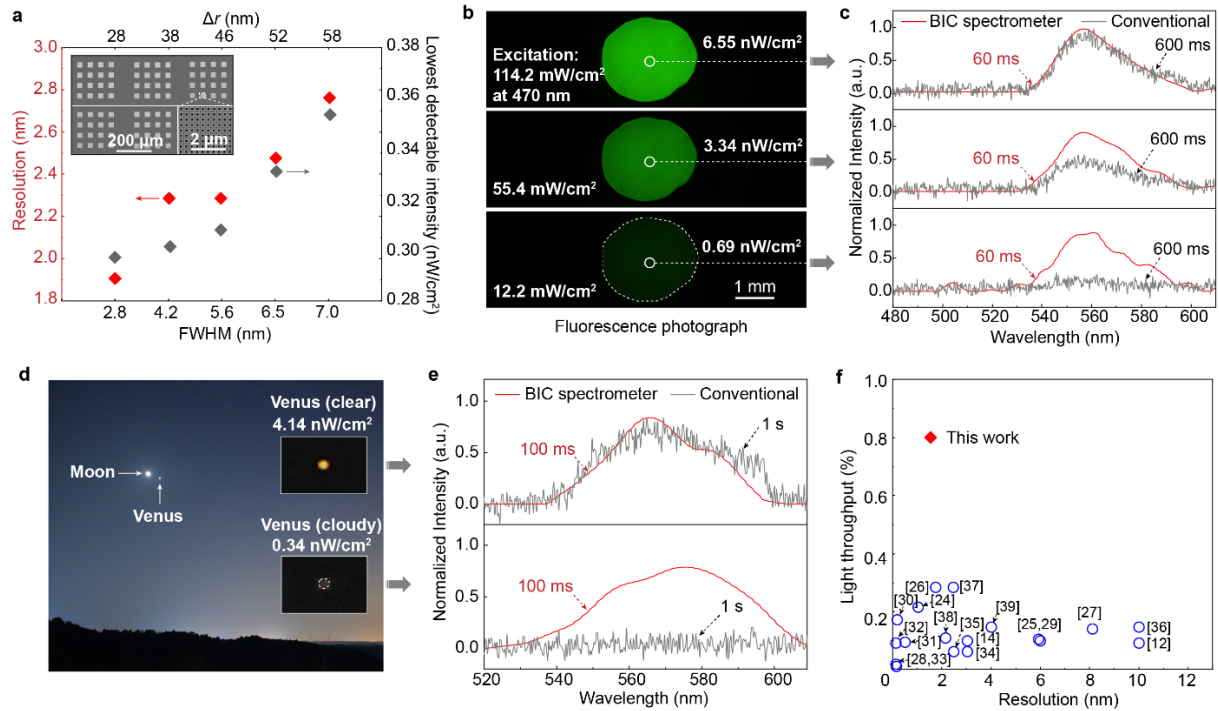

**Fig. 4 | Demonstration of the metasurface spectrometer for low light level detection. a**, Resolution and lowest detectable intensity of 5 qBIC-based spectrometers with different values of Δ$r$, each composed of 16 metasurface filter units, demonstrating a positive correlation between resolution and sensitivity. Inset shows SEM images of the 5 different spectrometers as well as a magnification of one filter unit (bottom right). **b,** Fluorescence photographs of bacteria under different excitation intensities and **c,** corresponding spectra measured by a qBIC-based metasurface spectrometer (red) and conventional mini-spectrometer (black), respectively. **d,** RGB photograph of the night sky, with magnified insets showing Venus in clear and cloudy conditions. **e,** Visible range spectra of Venus corresponding to clear (top) and cloudy (bottom) conditions from **d**, as measured by our BIC spectrometer versus a conventional mini-spectrometer. For both **c** and **e**, the conventional mini-spectrometer model is an Avaspec-uls2048cl (600/mm grating, 25 μm slit, 375000 counts/μW per ms integration time, which represents high-end commercial mini/micro-spectrometers). **f,** The light throughput efficiency (expressed as eventual intensity onto the detectors as a percentage of total light intensity before spectral dispersion / selection) and resolution of a selection of state-of-the-art (reconstructive and conventional) miniaturized spectrometers[12,14,24-39].

As discussed previously, the fundamental advantage of this qBIC based system is that, due to the nature of the spectral responses, the sensitivity should actually increase with resolution, rather than exhibiting a resolution-luminosity trade-off. As shown in Fig. 4a, to study the relationship between the resolution and sensitivity in our qBIC-based platform, five metasurface spectrometers were fabricated with different bandstop line widths and their performance was studied with respect to resolution and the minimum detectable irradiance, as a measure of their sensitivity. The spectrometers contain 16 metasurface filters with $\Delta r$ = 28, 38, 46, 52 and 58 nm, corresponding to FWHMs of 2.8, 4.2, 5.6, 6.5 and 7.0 nm, respectively. For a controlled comparison, in order to make these five spectrometers work over the same spectral range, the lattice constants, $P$, of each array element covers 336 - 351 nm, 333 - 348 nm, 330 - 345 nm, 326 - 341 nm, and 321 - 336 nm, respectively, in steps of 1 nm. These results shown in Fig. 4a suggest firstly that (for a constant number of filters) a smaller linewidth does indeed lead to higher spectral resolution, confirming our simulated work in the supplementary text (See Supplementary Information Section 1). We note that the maximum resolution of 1.9 nm achieved here is constrained by the number of filter array elements used (in this case 16 filter elements); the ultimate resolution limit for this device could be much lower. More crucially, the minimum detectable irradiance increased in correlation with the measured resolution. This supports our assertion that these systems are not constrained by a resolution-luminosity trade-off and can be engineered to achieve high sensitivity and spectral resolution, simultaneously, unlike in equivalent narrowband filter array systems. Through direct comparison with an equivalent reconstructive device, we further confirm the advantages of our platform over narrowband systems, in enhanced sensitivity through a greater utilization of available flux onto the detector (See Supplementary Information Section 4).

**Performance under ultra-low irradiance**

The key strength of this platform is in providing high sensitivity so as to maximise the limited available light in highly miniaturized systems. We first demonstrate the real-world applicability of the qBICs metasurface spectrometer's sensitivity via fluorescence measurements of bacteria samples (See Methods). The samples are characterized under three different illumination intensities, as shown in Fig. 4b. Comparisons between spectra measured by the qBICs metasurface spectrometer and a conventional mini-spectrometer are shown in Fig. 4c. As the excitation power decreases, the spectra measured by the conventional spectrometer becomes overridden by noise, while the qBICs metasurface spectrometer can still reconstruct the fluorescence spectrum. Enabling low-light excitation fluorescence measurements could help to avoid damage or photobleaching when measuring biological samples. Further to this, as shown in Fig. 4d, we use our platform to gather spectral data from the night sky, through telescopic measurement of Venus (See Methods). We show that even for clear conditions, where measurements become noise-dominated in a conventional device at an integration time of 1 s, our system produces clear data with order of magnitude lower integration times (Fig. 4e). We note that the irradiance for both fluorescence and planetary demonstrations cannot be directly measured by a power meter; we estimate the stated values *via* the linear power response of the qBIC spectrometer (See Supplementary Information Section 5). Furthermore, when cloudy conditions make measurement of the conventional system untenable, our device is still capable of collecting useful spectral data. We note here that through tuning the geometric parameters of the qBIC array the operational range could be straightforwardly extended into the infrared (See Extended Data Fig. 8), which is of more relevance to many astrophotonic applications. Finally, in Fig. 4f, we compare our work with other miniaturized spectrometer systems, based on both reconstructive/computational and direct-readout systems. As can be seen, the BIC system in this work conveys significant advantages with respect to photonic efficiency over, while maintaining a similar resolution to the state-of-the-art devices.

**Outlook**

We have demonstrated that by engineering metasurfaces with qBICs, it is possible to produce the building blocks of a new paradigm of micro-spectrometers that are not bound by the same limits on sensitivity that constrain conventional filter array systems. Augmented by a computational spectral reconstruction algorithm, these devices offer a route toward a platform that maintains high resolution and sensitivity in a highly miniaturizable, planar form. Such devices could prove ideal for a wide range of low-light-level applications such as Raman measurements, astronomical spectrographs and nanoscale biomedical spectroscopy.

**Methods**

**Device fabrication.** The fabrication of our metasurface consists of two main steps: Deposition of a $TiO_2$ layer, and the production of nanoholes in the $TiO_2$ film. $TiO_2$ films were deposited onto a quartz substrate via electron-beam evaporation. Substrates were first cleaned through successive ultrasonication steps in acetone, methanol, and deionized water, respectively, for 20 minutes each before electron-beam evaporation at a base vacuum of $2\times10^{-7}$ Torr and a deposition rate of 0.8 Å/s. After $TiO_2$ film deposition, absorptive coefficient, $k$, and refractive index, $n$, are characterized through ellipsometry. As illustrated in Extended Data Fig. 1, the absorption coefficient vanishes within the wavelength range of 480 to 610 nm, while the refractive index is approximately 2.3. As summarized in Extended Data Fig. 2, electron beam lithography (EBL) and Ion Beam Etching (IBE) processes were used to fabricate nanoholes in the films ensuring high accuracy and precision.

A three-stage mask process was used, with an EBL patterned resist to mediate etching onto a Cr:Au layer, which in turn was used to etch an amorphous silicon (a-Si) film that acted as the final mask for the $TiO_2$. A 200 nm a-Si layer was first deposited by plasma-enhanced chemical vapor deposition (SYSTEM 100) on the $TiO_2$ film. This was followed by a 5 nm Chromium (Cr) adhesion layer and a 75 nm Gold (Au) film, deposited by E-beam evaporation (COOKE). A 300 nm photoresist (ARP) film was then spin-coated onto the chip and baked at 85°C for 90s.

The designed nanohole lattice was patterned into ARP by an E-beam writer (MA6, with an acceleration voltage of 30 kV), before development in REX3038 solution and deionized water at 25 °C for 60 s and 30 s, respectively. The Cr:Au mask layer was then etched using the IBE, before etching of the a-Si layer through an Inductively Coupled Plasma (ICP) process. The ARP and Cr:Au mask layers were then removed by an oxygen plasma and wet chemical etch, respectively. Finally, the nanoholes were patterned into the $TiO_2$ layer by IBE through the Si mask, before removal of the mask by another ICP etch.

The choice to use three masks (ARP, Cr:Au, a-Si) was made as the $TiO_2$-etching gas readily etches photoresist and thus a hard mask is needed to pattern the $TiO_2$. Amorphous silicon was chosen as the hard mask material to avoid any influence on the optical properties of the BIC layer from metallic residues. However, to obtain an a-Si mask with accurately defined nanohole dimensions, the mask used to pattern it must be very thin (< 100 nm); a resist mask of this thickness would be removed too quickly during the a-Si etch, so an Au mask is used instead, which itself can be first patterned via the EBL resist.

**Numerical analysis of D-qBIC1 and D-qBIC2.** The studied metasurface can support two sets of D-qBICs in the frequency range of interest. The D-qBIC1 (D-qBIC2) are transverse-electric-like (transverse-magnetic-like) polarized, whose electric fields in the middle plane are out of the plane (in the plane). In the main text, we mainly focus on the D-qBIC1, as discussed in Fig. 2. Here, we show the magnetic field distribution of D-qBIC2 and their Q factor dependence on $\Delta r$, as illustrated in Extended Data Fig. 3. Compared with D-qBIC1, D-qBIC2 has a much larger Q factor and a much narrow FWHM in the case of the same $\Delta r$. As a result, the behaviour of D-qBIC1 is by far the dominant factor in the operation of the spectrometer (See Supplementary Information Section 3).

**Radiation properties of D-BICs and D-qBICs.** Simulated electric field distributions of D-BIC1 and D-qBIC1 in the yoz plane are displayed in Extended Data Fig. 4. It is seen that the field of D-BIC1 is strongly confined near the metasurface along the normal direction without any escaping radiation, implying an infinite Q factor. However, for D-qBIC1, we can observe that there is radiation out of the metasurface body, although the field is still localized along the normal direction of the metasurface, which makes qBIC an ideal candidate for optical spectrometers breaking the resolution-luminosity limit.

**Calibration.** To calibrate the spectral characteristics of the proposed spectrometer, a homemade calibration setup was constructed. Due to the narrow FWHM of BICs, it is necessary to find a monochromatic light source for calibration that has a narrower bandwidth, and, covering the operational range of our device, can be tuned in the wavelength region of 440 nm ~ 620 nm. As illustrated in Extended Data Fig. 5a, the light source used was a Xenon lamp (Microsolar 300), providing a white light source; a monochromator (CME-Mo301) was then used to disperse the light spatially, and monochromatic light of wavelength is selected by tuning the position of a grating. Here, a bandpass filter is also used to remove the harmonic wave stemming from the grating in the monochromator. The FWHM of the output light is 0.5 nm. The calibration light is divided by a 50:50 splitter, with the two beams directed to the spectrometer chip and a commercial spectrometer, respectively. The commercial spectrometer is used to monitor the central wavelength and power of the calibration light. The spectrometer chip is realized by aligning and attaching the metasurface onto the sensor surface of a CMOS pixel array. The CMOS chip is from *Indigo Inc.* (model: S400MRU), with a pixel size of 6.5µm $\times$ 6.5µm. The commercial spectrometer is an *Avaspec-uls2048cl* which operates with a resolution of 0.5 nm. Using the calibration setup, the spectral responses of the meta-spectrometer are characterized, as shown in Extended Data Fig. 5b.; here it can be seen clearly that the bandstop peaks are linearly dependent on the P value (i.e. pitch) of the nanohole lattice.

**Reconstruction algorithms.** The photoelectric response of the spectrometers' sensors can be expressed in the form of the integral equation (Eq. 1). For any set of response functions $T_i(\lambda)$ that are continuous over a bounded domain of $\Omega := [\lambda_1, \lambda_2]$, the reconstruction problem is inherently ill-posed, where the presence of a small perturbation in the response signal $I_i$ could produce an infinitely large change in the reconstructed spectrum[40]. Physically, such ill-posedness reflects the unavoidable information loss during the encoding procedure (Eq. 1), making the spectrum $S(\lambda)$ impossible to be directly obtained from the measurements $I_i$. If the incident spectrum $S(\lambda)$ can be well approximated by a function $\hat{S}(\lambda)$ that lies in the function space spanned by basis functions $\varphi_j$ with $j = 1, 2, \ldots, n$, then we can discretize the problem by the expansion:

$$S(\lambda) \approx \hat{S}(\lambda) = \sum_{j=1}^{n} \gamma_j \, \varphi_j(\lambda) \tag{2}$$

where the coefficients vector $x = [\gamma_1, \gamma_2, \ldots, \gamma_n]^\mathsf{T}$. In our case, we discretized the spectrum $S(\lambda)$ using piecewise polynomial and approximated the integral by *Gaussian Quadrature*[41]. Compared with Gaussian basis functions used in our previous studies[42], piecewise polynomials are more expressive and eliminate the need to manually select basis function parameters (i.e. FWHM of Gaussian bases), while the use of *Gaussian Quadrature* ensures the *exactness* of numerical integration as long as the integrand is a polynomial with a fixed degree[41].

Then, problem (Eq. 1) can be transformed into a finite-dimensional least-square problem in the form of:

$$x_{LS} := \min_{x} \frac{1}{2} \| Ax - b \|_2^2 \tag{3}$$

where $b := [I_1, I_2, \ldots, I_m]^\mathsf{T} \in \mathbb{R}_+^m$ is the vector of photoelectric responses and $A$ is the coefficient matrix with each element $a_{ij}$ determined by the inner product of response function $T_i$ and basis function $\varphi_j$ over the bounded domain $\Omega$.

To regularize problem (Eq. 3), we imposed both $l_1$ and $l_2$ regularization alongside with the non-negative constraint on $x$:

$$x^\dagger_{\alpha,\beta} := \min_{x \geq 0} \frac{1}{2} \| Ax - b \|_2^2 + \alpha \| x \|_1 + + \frac{1}{2}\beta^2 \| x \|_2^2 \tag{4}$$

where $L \in \mathbb{R}^{(n-1)\times n}$ is a first-order discrete gradient matrix that constrains the smoothness of the reconstructed solution. By using the non-negative constraint of $x$, the minimization problem (Eq. 4) can be expanded into a constrained quadratic programming problem:

$$x^\dagger_{\alpha,\beta} := \min_{x \geq 0} \frac{1}{2} x^\top (A^\top A + \beta^2 L^\top L)x + x^\top (\alpha \mathbf{1}_n - A^\top b) \tag{5}$$

where $\mathbf{1}_n = [1,1,\ldots,1]^\top \in \mathbb{R}^n_+$.

The two regularization terms serve different purposes. The $l_1$ norm will promote sparsity which ensures high-resolution reconstruction for narrow spectra, while the $l_2$ norm regularization on the gradient will increase the smoothness of the reconstruction, thus more effective for broad spectra. The relative importance of sparsity and smoothness is measured by regularization parameters $\alpha$ and $\beta$ respectively, which are selected using *K-fold cross-validation*[43].

**Fluorescence and astronomical spectroscopy measurements.** To verify the high-throughput ability of the spectrometer chip, a weak fluorescence spectrum of bacteria is measured. The test setup for this experiment is shown in Extended Data Fig. 6. The LED source emits broadband light, from which the fluorescence-excitation is selected using a bandpass filter of 420 nm ~ 485 nm. A beam splitter (BS1) is used to reflect the light onto a fluorescent sample via a 4× microscope objective. The fluorescence and the reflected excitation light is collected by the same objective. A 515 nm longpass filter removes residual excitation light, leaving only the fluorescence signal. The fluorescence is divided equally by a 50:50 beam splitter (BS2). One beam is incident on the metasurface-spectrometer while the other one enters the commercial spectrometer. Adjustments are made so that the apertures of the commercial spectrometer and meta-spectrometer are conjugated and as such the fluorescence measured by the two spectrometers comes from the same point on the fluorescent sample.

The Venus measurement setup is shown in Extended Data Fig. 7a and b. Venus was photographed with an astronomical telescope CGEM-II-1100HD from Celestron Co., Ltd. The telescope has a Schmidt-Cassegrain configuration with an aperture of 11 inches and a focal length of 2800 mm. A lens with a 5× magnification is used to focus the light captured by the telescope onto the BIC spectrometer. A 50:50 beam splitter is placed in front of the BIC spectrometer to split half of the light into the conventional, grating-based spectrometer. The BIC spectrometer and the conventional spectrometer have an aperture with the same size. Between the imaging lens and the beam splitter, a 412 – 569 nm bandpass filter FGB18 from Thorlabs and a 550 nm longpass filter CB550 from Yongxing-Sensing Co. Ltd are mounted to filter the Venus light to those within the operating wavelength of the BIC spectrometer. As well as this, a 600 nm lowpass filter is also used to reduce noise from background infrared light. The clear image of Venus was taken at about 21:00 and the cloudy one was taken at about 21:40.

**Method references**

**Acknowledgments**

The authors acknowledge funding from National Natural Science Foundation of China (grants No. 62104212; 61705204; U20A20172), National Key R&D Program of China (grants No. 2022YFB3206000), China Postdoctoral Science Foundation (grants No. 2019M653486), Cao Guangbiao High Science and Technology Foundation, Zhejiang University (grants No. 2022RC011), Fundamental Research Funds for the Central Universities (grants No. 2021FZZX001-07), Zhejiang University Education Foundation Global Partnership Fund (grants No. 100000-11320), Natural Science Foundation of Sichuan Province (grants No. 2022NSFSC1728), Zhejiang Province Key R & D programs (grants No. 2021C05004).

The authors thank Dong Liu, Hao Wu, Feng Jing, Yuqiu Gu, Yinpeng Chen, Haijun Zhang, Xuan Luo, Rong Qiu, Chaoyang Wang, Zhixi Li, Zhiqiang Zhan, Tianchun Xu and Zehai Guan for discussion. We thank Weiyi Zhang and Jing Cui from Ideaoptics Co., Ltd and Qiang Li from Sichuan society for amateur astronomy for the experimental assistance. The authors also thank the Micro and Nano Fabrication Centre at Zhejiang University for facility support.

**Author contributions**

F.T., Z.Y.Y., X.Y. and Y.H.Y. initiated the project. H.X.C. and F.T. performed the calculation. F.J.C., Y.H.Y., Y.Q.S. and F.T. performed the simulation. Z.Y.Y., Y.H.Y. and T.A.O. designed the experiments. J.J.W. fabricated samples. F.T., J.J.W., L.Z., J.C., J.H., L.M.Y. and W.G.Z. carried out the measurements. Z.Y.Y., Y.H.Y., F.T., H.X.C., B.L., S.R.D., J.K.L., B.F.J., X.H.G., S.L.L., Y.J.S., L.F.C., Y.Q.S., H.S.C., T.A.O., T.H., E.A.M. and H.J.J. analyzed the results and wrote the manuscript. Z.Y.Y., T.H., X.Y. and Y.H.Y. supervised the project.

**Competing interests**

The authors declare no competing interests.

**Correspondence and requests for materials** should be addressed to Zongyin Yang.

**Data availability**

The data that support the plots within this paper are available from the corresponding authors upon request.

**Code availability**

The code used in this paper are available from the corresponding authors upon request.

**Extended data figures**

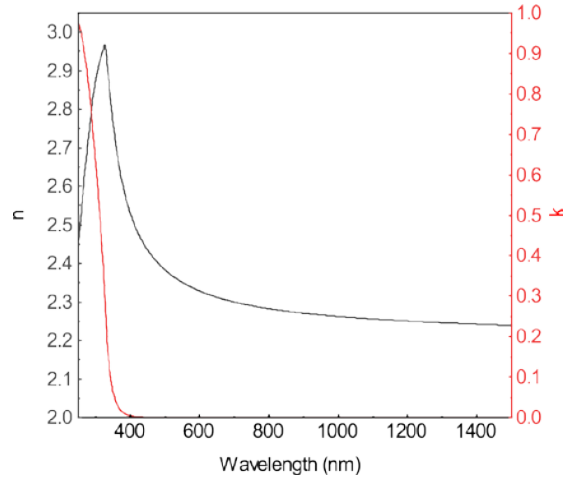

**Extended Data Fig. 1 | Optical properties of the TiO$_2$ film.** Showing refractive index, n (black) and absorptive coefficient, k (red).

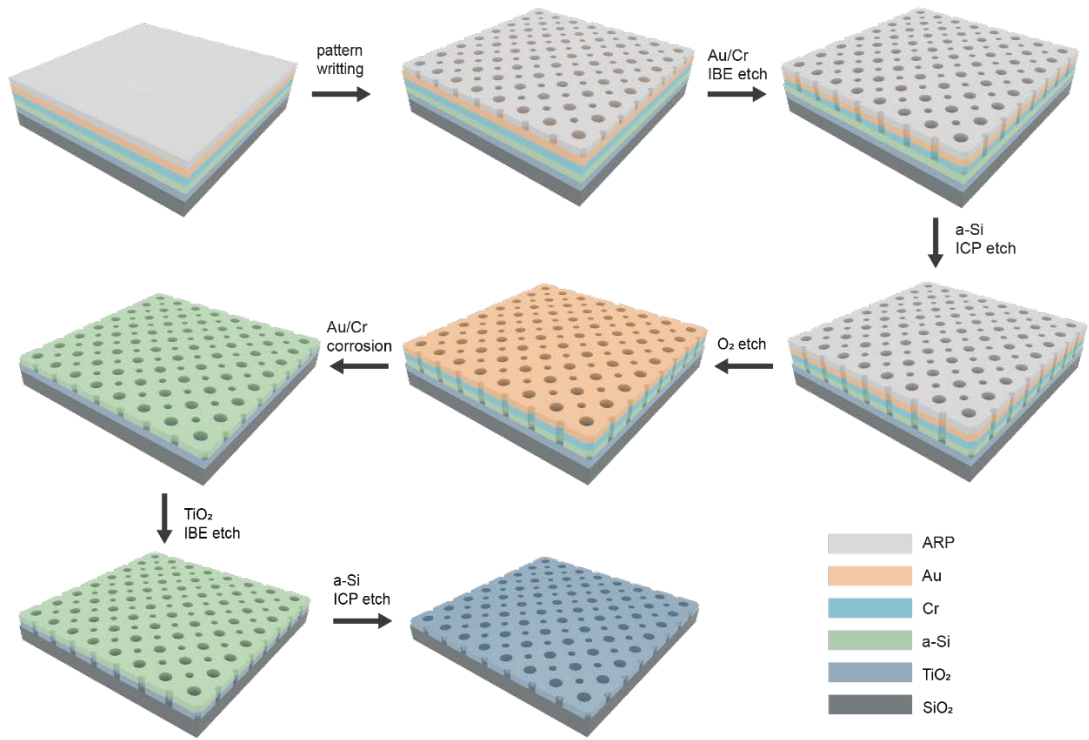

**Extended Data Fig. 2 |** Schematic showing the fabrication process of a nanohole lattice in the TiO$_2$ film to create a metasurface slab.

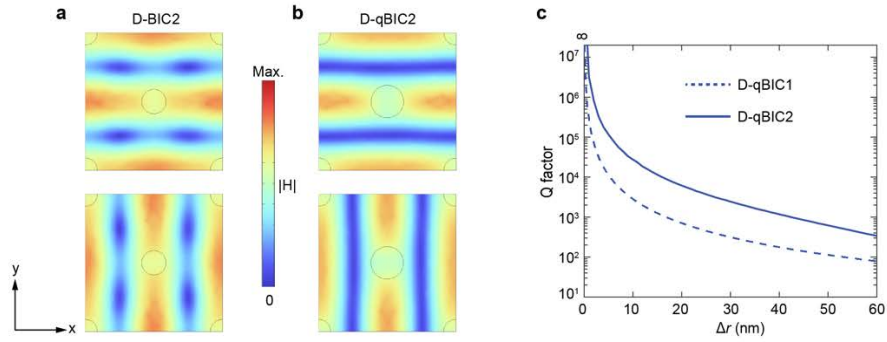

**Extended Data Fig. 3 | Numerical analysis of D-qBIC1, D-qBIC2 and D-qBIC2.** (a, b) Simulated Hz magnetic field distributions of D-BIC2 (a) and D-qBIC2 (b) at the z=0 nm xoy plane. (c) Q factor of D-qBIC1 (dashed curve) and D-qBIC2 (solid curve) extracted from simulated transmitted spectra as a function of Δr. Note that as these two modes are doubly degenerated, they have the same Q factor.

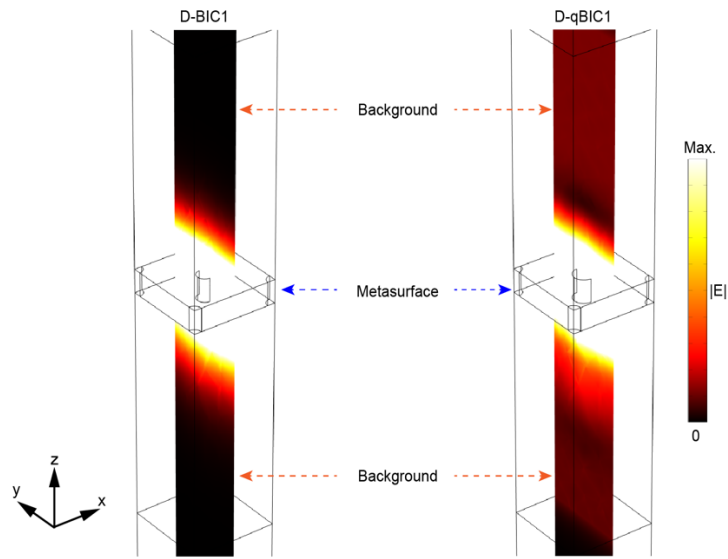

**Extended Data Fig. 4 | Radiation properties of D-BIC and D-qBIC.** Simulated electric field distributions of D-BIC1 and D-qBIC1 in the yoz plane. The D-BIC2 (D-qBIC2) have similar radiation characteristics to D-BIC1 (D-qBIC1).

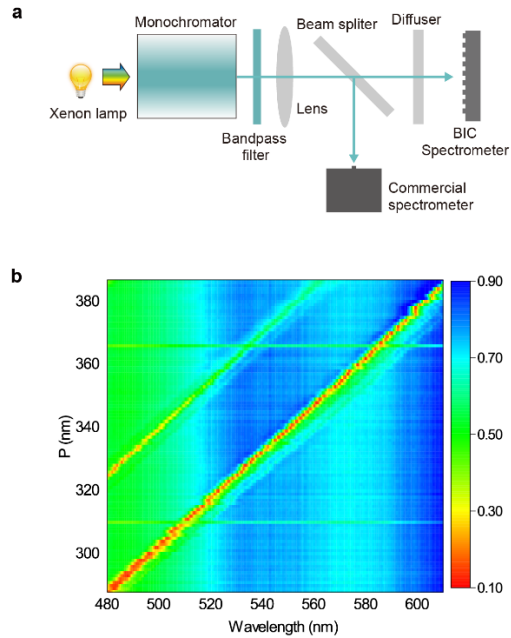

**Extended Data Fig. 5 | Reconstruction fidelity experimental setup. a,** Diagram showing the apparatus for characterizing the spectral features of the spectrometer chip. **b,** Spectral response profiles of the spectrometer as a whole. The response of the spectrometer is a combination of spectral transmission of BIC filters and the response of the CMOS sensors beneath.

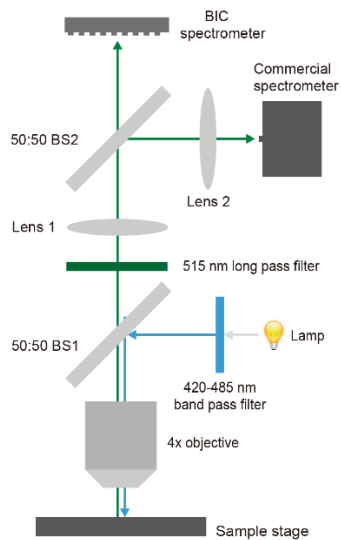

**Extended Data Fig. 6 |** Diagram showing the experimental apparatus for measuring the fluorescence of a bacterium under low light conditions.

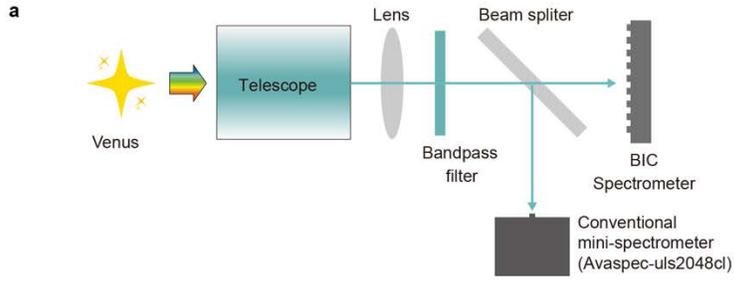

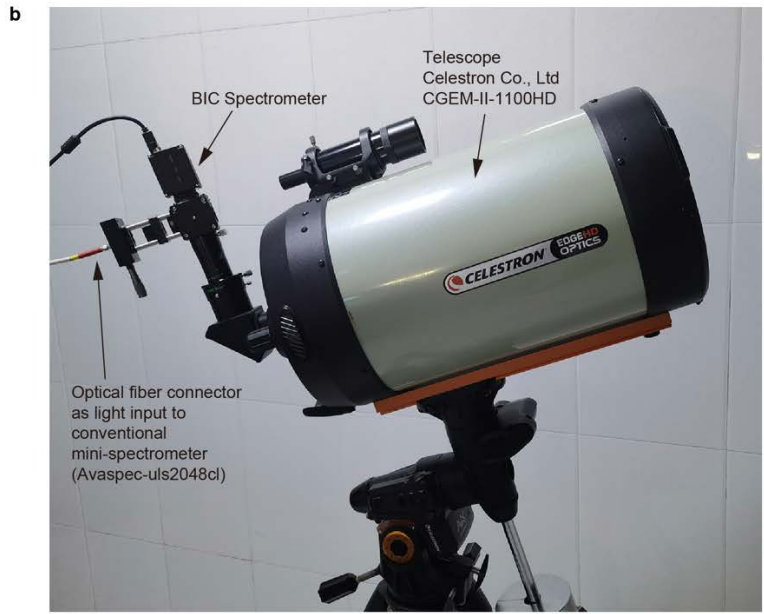

**Extended Data Fig. 7 | a,** Schematic and **b,** picture of the setup for measuring the spectrum of Venus.

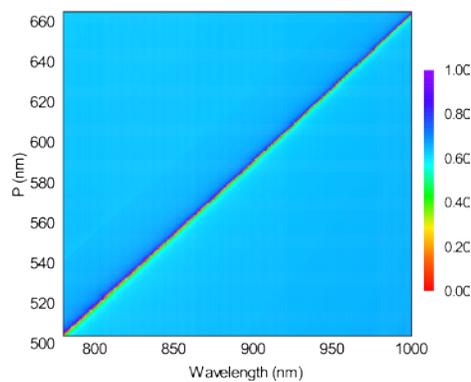

**Extended Data Fig. 8 |** Extension of BIC spectral positions to near infrared region via increasing the pitch value P.

**End**